\documentclass[11pt]{article}         

\begin{document}




\title{Nonsingularity of Flat Robertson-Walker Models in the Special Relativistic Approach to Einstein's Equations}
\author{J. Brian Pitts\footnote{
Mathematics Department, 
	St. Edward's University,
	Austin, Texas 78704; and
	The Ilya Prigogine Center for Studies in
     Statistical Mechanics and Complex Systems, 
	Department of Physics,
     The University of Texas at Austin,
     Austin, Texas 78712;
	current address: Program in History and Philosophy of Science
				and Department of Philosophy,
				University of Notre Dame,
				Notre Dame, Indiana 46556;
	email jpitts@nd.edu
\vspace{2 mm} }
  and W. C. Schieve\footnote{
The Ilya Prigogine Center for Studies in
     Statistical Mechanics and Complex Systems, 
	Department of Physics,
     The University of Texas at Austin,
     Austin, Texas 78712} } 

\date{\today}

\maketitle


\begin{abstract}

     Recently the neglected issue of the causal structure in the flat spacetime approach to Einstein's theory of gravity has been substantially resolved.  Consistency requires that the flat metric's null cone be respected by the null cone of the effective curved metric.  While consistency is not automatic, thoughtful use of the naive gauge freedom resolves the problem.  After briefly recapitulating how consistent causality is achieved, we consider the flat Robertson-Walker Big Bang model. The Big Bang singularity in the spatially flat Robertson-Walker cosmological model is banished to past infinity in Minkowski spacetime.  A modified notion of singularity is proposed to fit the special relativistic approach, so that the Big Bang becomes nonsingular.

\end{abstract}


keywords:  bimetric, causality principle, null cones, cosmological singularity

\section{Introduction}

     A number of authors have discussed the possibility of deriving Einstein's equations from postulates of field theory in Minkowski spacetime \cite{SliBimGRG,NullCones1} (and references therein).  Such derivations are quite simple and compelling.  The governing intuition is to formulate gravitation as much like other forces as possible.  All other forces typically live in Minkowski spacetime, and must not accelerate objects so as to move outside the light cone.  Avoiding an acausal theory of gravity in Minkowski spacetime implies that the null cone of the effective curved metric $g_{\mu\nu}$ must respect the null cone of the flat background metric $\eta_{\mu\nu}$.  We call this condition $\eta$-causality.  These matters have been studied in detail elsewhere \cite{NullCones1}.  While such consistency between the two null cones does not obtain automatically, thoughtful use of the naive gauge freedom of the theory plausibly permits the correct relation between the two null cones.  Given a change of variables, the configuration space of the theory is reduced to essentially just those curved metrics that respect $\eta$-causality.  The smaller configuration space entails that gauge transformations, which must connect only curved metrics within the smaller configuration space, form a groupoid, not a group, because the binary operation is not defined for every pair of elements.  The relation between the two null cones is now both physically satisfactory and gauge invariant.  

	The use of Minkowski spacetime as the stage for gravitational phenomena has two obvious benefits.  First, the flat background metric's null cone provides a causal structure for defining dynamics in quantum gravity.  Second, it also ensures that the curved metric will describe a globally hyperbolic spacetime.  In this way, some of the difficulties of the geometrical theory are removed naturally.

	These benefits follow from the conformal part of the flat metric.  The primacy of Minkowski spacetime, however, implies that we must consider its full metric structure, not merely its causal structure.  This metric structure also has physical consequences.  Given that Minkowski spacetime is the stage for all phenomena, it follows that all solutions of Einstein's equations admissible in the special relativistic approach (SRA) must exactly fit onto Minkowski spacetime, while respecting $\eta$-causality.  In particular, there must not be regions of spacetime that simply lack a curved metric $g_{\mu\nu}$.  

	In parallel with standard requirements in the geometrical approach, one wishes to extend the solution of Einstein's equations as far as possible.  Then the range of affine parameter for $g$-geodesics is as great as Minkowski spacetime permits.  Let us call this principle ``maximal SRA extension,'' because it modifies the general relativistic desideratum of maximal extension \cite{Wald} to ensure consistency with the special relativistic approach.   This principle prevents the gratuitous truncation of solutions.
	
	Here we consider the flat Robertson-Walker Big Bang model.  This solution has a finite past temporal extent, as measured using $g_{\mu\nu}$.  Minkowski spacetime, on the other hand, has infinite past temporal extent, as measured using $\eta_{\mu\nu}$.  \emph{Prima facie} it is not obvious whether a physically reasonable way to map the Robertson-Walker model into Minkowski spacetime even exists.   But one certainly hopes so, given the cosmological importance of the Robertson-Walker solution.  We shall see that a physically reasonable mapping indeed does exist.  It turns out that the Big Bang singularity is naturally mapped into the infinite past as determined by $\eta_{\mu\nu}$.  The singularity is therefore abolished, because one does regard misbehavior at infinity as a singularity.  It follows that standard technical definitions of singularity in general relativity, which do not refer to a background metric, do not suit the special relativistic approach to Einstein's equations.  We sketch how the notion of singularity might be modified.  
	
 \section{Gauge Transformations}

	In any physical theory, a  gauge transformation changes the field variables in such a way that the action is changed at most by a boundary term.  The Euler-Lagrange equations are thereby preserved (assuming that no worries about the domain occur).  For a simple theory like Maxwell's electromagnetism, one can easily distinguish gauge effects from physics by directing one's attention from the vector potential to the electromagnetic field strength.  For Einstein's equations with a flat background metric, a transformation of the form 
$g_{\mu \nu }\rightarrow e^{\pounds _{\xi}}g_{\mu\nu},
\eta _{\mu \nu}\rightarrow \eta _{\mu\nu},
u\rightarrow e^{\pounds _{\xi }}u $ changes the action by at most a boundary term, and so is a candidate for being a gauge transformation.  However, physical equivalence requires more than just changing the action by a boundary term, because there is more to physics than field equations \cite{SorkinScalar}.  In particular, one must specify the topology and the boundary conditions, as Sorkin notes.  In a bimetric theory, one must also specify what causality means.

Given the causal structure of Minkowski spacetime as a bound on the curved metric, two solutions of Einstein's equations can only be regarded as equivalent if they have qualitatively the same relationship between the two null cones.  It follows that only certain vector fields generate gauge transformations. Furthermore, the nonlocal form due to infinite order derivatives in the gauge transformation formula above implies that the vector field $\xi^{\mu}$ must obey some boundary conditions, so that no piece of the solution of the field equations is added or omitted by the transformation.  Vector fields not respecting the causal structure and the boundary conditions do not generate gauge transformations: although they change the action by only a boundary term, they nevertheless relate physically inequivalent states.  To be precise, a gauge transformation must be redefined as an ordered triple of the form 
\[
( e^{\pounds_{\xi} }, \eta_{\mu\nu}, g_{\mu\nu}),
\]
but with the requirements that both the original and the transformed curved metrics 
respect  $\eta$-causality, and that $\xi^{\mu}$ goes to 0 fast enough at the boundaries.  

	One wants to compose two gauge transformations to get a third gauge transformation.  At this point, the fact that a gauge transformation is not labelled merely by the vector field, but also by the curved and flat metrics, has important consequences.  Clearly the two gauge transformations to be composed must have the second one start with the curved metric with which the first one stops. 
 We also want the flat metrics to agree.  Thus, the `group' multiplication operation is defined only in certain cases, meaning the gauge transformations in the special relativistic approach \emph{do not form a group}, despite the inheritance of the mathematical form of exponentiating the Lie differentiation operator from the field formulation's gauge transformation. Instead they form a groupoid \cite{NullCones1}, as defined by P. Hahn \cite{Hahn} and J. Renault \cite{Renault}.  According to A. Ramsay, ``[a] groupoid is, roughly speaking, a set with a not everywhere defined binary operation, which would be a group if the operation were defined everywhere.'' \cite{Ramsay} (pp. 254, 255)  The physics requires redefining gauge transformations in this way.  As a consequence, the relationship between the two null cones is now gauge invariant, as anything physically significant must be.  
	

\section{Exiling the Big Bang to the Infinite Past in the Flat Robertson-Walker Model}

	As we will soon show, the Big Bang singularity for the flat Robertson-Walker cosmological spacetimes is dissolved by exile to the infinite past.  (This idea formally resembles a suggestion made by C. Misner \cite{MisnerTime}, A. G. Agnese and A. Wataghin \cite{Agnese}, 
and J.-M. Levy-Leblond \cite{Levy-Leblond} in general relativity, but with a different justification.)  To make this fact clear, let us recall key features of the  special relativistic approach to Einstein's equations.  First, the special relativistic approach, having two metrics, has two notions of length and time. ``Unrenormalized'' lengths and times are determined by the flat metric $\eta_{\mu\nu}$.  ``Renormalized'' lengths and times are measured by real rods and clocks, which are distorted by the gravitational field and so are governed by $g_{\mu\nu}$.  Let us use $s$ for renormalized proper time and $\sigma$ for unrenormalized time, both for a fundamental observer in a flat Robertson-Walker model.  Observable features of the world specify the dependence of the scale factor $a$ on $s$, but not the dependence of $a$ on $\sigma$, or, equivalently, of $s$ on $\sigma$.  Robertson-Walker cosmological solutions to Einstein's equations imply a past of finite renormalized duration, but do not obviously say anything about the unrenormalized duration.  Second, all solutions in the special relativistic approach must fit onto Minkowski spacetime, leaving no bare patches, while respecting $\eta$-causality.  Third, such solutions must be maximally SRA extended.  

	While curved geometry admits three qualitatively distinct homogeneous and isotropic Robertson-Walker models, installing the curved cases into Minkowski spacetime is not straightforward.  For the negatively curved case, an inhomogeneous relation between the two metrics might be required.  The positively curved case, we suspect, cannot be included at all. Therefore we consider only the spatially flat Robertson-Walker model here.  
	
	Clearly the renormalized proper time of the universe $s$ must be a strictly increasing function of the unrenormalized proper time $\sigma$, but considerable arbitrariness  remains in their functional relationship.  (Fixing a relation amounts to partial gauge-fixing.)  The possibilities that one might naively entertain fall into three classes.   We can readily express them using the dependence of the scale factor $a$ on $\sigma$ (not $s$).  First, $a(\sigma)$ might go to $0$ at some finite value of $\sigma$, so that the Big Bang happened in the finite unrenormalized past.  Second, $a(\sigma)$ might exceed some fixed $\epsilon > 0$ for all $\sigma > -\infty$, so that the Big Bang simply fails to occur, the scale factor being finite even at arbitrarily early moments  of $\sigma$.  Third, $a(\sigma)$ might go to $0$ as $\sigma \rightarrow -\infty$, so that the Big Bang `happened' only in the limit of the infinite unrenormalized past.  We will consider each of these three options.  Finding two of them unsatisfactory, the remaining  one must be correct.  

	 The first option implies that the early end of Minkowski spacetime would be simply devoid of any state of the gravitational field at all.  This is not to say that the gravitational field vanished, but that there simply was no fact of the matter about the gravitational field, throughout an infinite past.  Then one Sunday evening, \emph{according to the laws of physics} (with no special act of God), a genuine physical singularity of infinite curvature arose everywhere at once, for no reason.  From this conflagration emerges a universe, and the usual Robertson-Walker history follows.  After waiting quietly forever, sheer nothingness in Minkowski spacetime spontaneously became singular and then evolved into all things.  But such a story is repugnant to reason. Out of nothing, nothing comes, especially in classical physics, so this option can be discarded.

	Turning to the second option, we avoid the previous absurdity, but a new difficulty arises.  The second option violates maximal SRA extension by gratuitously truncating the solution of Einstein's equations.  The mathematics permits the existence of scale factors that are arbitrarily close to 0, but this second option says that some of them simply were never instantiated in an infinite past.   Moreover, there exists some greatest lower bound $\epsilon$ for the scale factor.  What could explain the value of $\epsilon$?  Absent some justification for violating maximal SRA extension, we reject this second option as implausible. 

	 The only remaining alternative is that the scale factor goes to $0$ as  $\sigma \rightarrow -\infty$, in the unrenormalized infinite past.  This alternative obeys maximal SRA extension, and yet avoids leaving part of Minkowski spacetime bare, unlike the other two options.  It is clear, then, that in the special relativistic approach, the Big Bang `happened' in the unrenormalized infinite past.  The finite renormalized time that elapsed along the same worldline shows that arbitrarily large gravitational time dilation occurred in the early universe, with the curved metric's lapse function approaching $0$, in the flat Robertson-Walker model.

  Is it fair to say that the Big Bang singularity is thereby removed?  In the context of the geometrical theory based on Einstein's equations, Wald writes, ``even if the value of some curvature scalar is unbounded in a spacetime, the curvature might blow up only `as one goes to infinity,' in which case one would not want to say that the spacetime is singular'' \cite{Wald} (p. 214), an intuition shared by other authors.  Wald lacked the distinction between renormalized and unrenormalized times and lengths that comes with two metrics, so our technical realization of this intuition will differ from his.  In the special relativistic approach, it is clear that the phrase ``as one goes to infinity'' must be cashed out in terms of \emph{unrenormalized} times and lengths, because these correspond to the outer reaches of Minkowski spacetime, the stage for all events.  A distance or duration can be $\eta$-infinite, but $g$-finite, simply due to arbitrarily strong gravitational distortion of rods or clocks.  Therefore, for the special relativistic approach, one must modify the suggestion \cite{Wald} (p. 215) that a singularity exists just in case there is a $g$-geodesic curve which is inextendible in at least one direction, but of finite $g$-length.  We must add the qualification that the curve in question have finite $\eta$-length.  

	Let us attempt to redefine singularities by making the smallest change possible in the definition above.  We suggest the following:  a gravitational field configuration is singular just in case there exists a  $g$-geodesic curve which is inextendible in at least one direction, but of finite $g$-length \emph{and} finite $\eta$-length.  (For curves that are null with respect to some metric,  one should read ``finite length'' in terms of an affine parameter corresponding to that metric.) With this definition of singularities, it is clear that the Big Bang at past unrenormalized infinity is not a singularity.  Thus the special relativistic approach removes the Big Bang singularity by exile to infinity. 
 
	We have not addressed spatially curved Robertson-Walker models.  One suspects that  positively curved homogeneous models simply are not possible in the special relativistic approach.  If so, then one predicts that either the net density of the universe is no greater than the critical value, or that homogeneity fails in some sense, either observably, or in the relation between the two metrics.  The inclusion of negatively curved models in Minkowski spacetime might require weakening the homogeneity condition in some way, perhaps rather subtle. We hope to address this issue soon.   It would be interesting to know if the modified notion of singularity, which takes note of the possibility of exiling singularities to infinity, removes  singularities from any other solutions of Einstein's equations besides the flat Robertson-Walker model.

\section{Conclusion}

	We have taken special relativity seriously, including its causal and metric structure, while viewing gravity as described by Einstein's field equations.  With a few innovations, one can arrange for the correct relationship between the two null cones to hold.  Taking Minkowski spacetime seriously yields physical benefits.  The causal structure of the flat metric defines the equal times in equal time commutation relations for quantum gravity, and guarantees global hyperbolicity.  The full Minkowski metric structure makes itself known by banishing the Big Bang to the infinite past in flat Robertson-Walker models.  Given a redefinition of singularity, the Big Bang singularity  disappears.




\begin{thebibliography}{10}


\bibitem{SliBimGRG} J. B. Pitts and W. C. Schieve, \emph{Gen. Rel. Grav.} {\bf 33}, 1319 (2001); gr-qc/0101058. 

\bibitem{NullCones1} J. B. Pitts and W. C. Schieve, ``Null cones in Lorentz-covariant general relativity,'' gr-qc/0111004 (2002).
%

\bibitem{Wald} R. Wald, \emph{General Relativity} (Univ. Chicago, Chicago,  1984).

\bibitem{SorkinScalar} R. Sorkin, ``An example relevant to the Kretschmann-Einstein debate'', philsci-archive.pitt.edu (2002).
%

\bibitem{Hahn}  P. Hahn, \emph{Trans. Amer. Math. Soc.} {\bf 242}, 1 (1978). %

\bibitem{Renault}  J. Renault, \emph{A Groupoid Approach to $C^{*}$-algebras} (Springer, Berlin, 1980). 

\bibitem{Ramsay} A. Ramsay, \emph{Advances in Mathematics} {\bf 6}, 253 (1971).

\bibitem{MisnerTime} C. W. Misner, \emph{Phys. Rev. A} {\bf 186}, 1328 (1969).  

\bibitem{Agnese} A. G. Agnese and A. Wataghin, \emph{Lett. Nuov. Cim.} {\bf 1}, 857 (1971). 

\bibitem{Levy-Leblond} J.-M. Levy-Leblond, 
\emph{Amer. J. Phys.} {\bf 58}, 156 (1990). 


\end{thebibliography}
\end{document}